\documentclass{svproc}
\usepackage{url}

\usepackage{amssymb}
\usepackage{amsmath}
\usepackage{mathtools}
\usepackage{xcolor}

\begin{document}
\mainmatter              
\title{Biarchetype analysis for univariate functional data. An application to macroeconomic financial time series}
\titlerunning{Biarchetype analysis for functional data}  
%
\author{Aleix Alcacer\inst{1} \and Rafael Ben\'itez\inst{2} \and
Vicente J. Bol\'os\inst{2}\thanks{Corresponding author} \and Irene Epifanio\inst{1}}
\authorrunning{Aleix Alcacer et al.} 
%
\tocauthor{Aleix Alcacer, Rafael Ben\'itez, 
Vicente J. Bol\'os and Irene Epifanio}
\institute{Jaume I University, Castell\'o 12071, Spain.\\
\and
University of Val\`{e}ncia,
Val\`{e}ncia 46022, Spain.\\
\email{vicente.bolos@uv.es}}

\maketitle              

\begin{abstract}
We introduce biarchetype analysis for the first time in the context of univariate functional data. This unsupervised methodology extends archetype analysis by simultaneously identifying archetypal structures across both the cases (countries, in our application) and the temporal argument. Both cases and time points are expressed as mixtures of biarchetypes, yielding a concise and highly interpretable representation of complex functional observations. Although biarchetype analysis is not intended as a clustering technique, it offers superior interpretability compared with biclustering approaches, as it is based on extreme, representative patterns rather than average centroids, thereby enhancing human comprehension. We apply the proposed method to 10-year government bond yields of European countries over the period 2001–2025. The results identify three distinct time regimes (the pre-crisis period, the euro-area sovereign debt crisis, and the post-crisis period), and reveal Germany, Greece, and Hungary as country archetypes. 
\keywords{archetype analysis, functional data analysis, bond yields, biclustering}
\end{abstract}
\section{Introduction}
Archetypal analysis (AA) \cite{cutler1994archetypal} is an unsupervised learning technique that identifies distinct patterns in high-dimensional data, representing each observation as a convex combination of these archetypes. The archetypes themselves are convex combinations of the data points, serving as idealized exemplars that capture the essential variation. This framework provides an interpretable approach to dimensionality reduction and feature extraction, as discussed in the survey \cite{alcacer2025survey}.

While standard AA identifies archetypes only among observations, \cite{palumbo2012archetypal} applied AA in the variable space to obtain archetypal variables. A recent extension, biarchetype analysis (biAA) \cite{alcacerieee24}, generalizes this concept by simultaneously identifying archetypes for both observations and variables, with each represented as a mixture of biarchetypes.  In other words, AA captures extreme patterns among observations, but biAA captures extremes jointly across both observations and variables. 
In this sense, biAA relates to AA as fuzzy biclustering relates to fuzzy clustering, and  biAA relates to fuzzy biclustering as AA relates to fuzzy clustering. 
Although biAA is not a clustering method, it can be employed to segment datasets, which is particularly useful when the data form a continuum. In \cite{alcacerieee24}, biAA was compared with several biclustering techniques, demonstrating clear advantages, especially with respect to interpretability.

biAA has thus far been applied only to multivariate data and has not yet been adapted for the analysis of functional data. In functional data analysis (FDA), the observations are functions rather than scalar values. An excellent overview of FDA is provided by \cite{Ramsay05}. AA was extended to dense functional data in \cite{Epifanio2016} and sparse functional data in \cite{VinEpi17,VinEpi19}.

In this work, we extend biAA to dense univariate functional data, representing its first application in this context. Applied to univariate functional data, biAA facilitates the joint identification of archetypes across both observations and the temporal line.
Section \ref{met} outlines the methodology, Section \ref{res} presents the data, macroeconomical financial time series, and results, and Section \ref{con} provides the conclusions.

\section{Methodology} \label{met}
Let $\{x_i(t)\}_{i=1}^n$ denote a sample of univariate functional observations defined on a common domain $\mathcal{T}$. In practice, these functions are observed on a grid of $m$ points, yielding the data matrix ${\bf X}_{n \times m}$.

In functional AA, the $k$ archetypes $\{z_g(t)\}_{g=1}^k$ are defined as convex combinations of the observed functions, $z_g(t) = \sum_{l=1}^n \beta_{gl} x_l(t)$,  
$\sum_{l=1}^n \beta_{gl} = 1$,  $\beta_{gl} \geq 0$,  $g=1,\ldots,k$,
which, in matrix form, corresponds to $\mathbf{Z} = \boldsymbol{\beta}\mathbf{X}$.

Each functional observation is then approximated as a convex combination of the archetypes,
$x_i(t) \approx \hat{x}_i(t)$ = $\sum_{g=1}^k \alpha_{ig} z_g(t)$, 
$\sum_{g=1}^k \alpha_{ig} = 1$,  $\alpha_{ig} \geq 0$, $i=1,\ldots,n$.

Under these constraints, the residual sum of squares to be minimized is
\begin{equation}
\label{eq:aa_fda}
RSS
= \sum_{i=1}^n \int_{\mathcal{T}}
\left(
x_i(t) - \sum_{g=1}^k \alpha_{ig} \sum_{l=1}^n \beta_{gl} x_l(t)
\right)^2 dt,
\end{equation}
which reduces to the discrete formulation $\left \Arrowvert {\bf X} - \boldsymbol{\alpha}\boldsymbol{\beta}\mathbf{X}\right \Arrowvert^2$ when the functions are evaluated on a finite grid,  where $\left \Arrowvert . \right \Arrowvert$ stands for the Frobenius norm.

We now extend this framework to functional biAA. Functional biAA simultaneously identifies biarchetypes in both the observation and the temporal dimensions. In functional biAA, the biarchetypes are defined as 
$\mathbf{Z}_{k \times c}$ = $\boldsymbol{\beta}_{k \times n}\, \mathbf{X}_{n \times m}\, \boldsymbol{\theta}_{m \times c}$, 
where $\boldsymbol{\beta}$ and $\boldsymbol{\theta}$ are convex combination matrices satisfying $\sum_{l=1}^n \beta_{gl} = 1$, $\beta_{gl} \geq 0$, $g=1,\ldots,k$, and
$\sum_{r=1}^m \theta_{rh} = 1$, $\theta_{rh} \geq 0$, $h=1,\ldots,c$. Thus, each biarchetype represents a convex combination of both observed functions and discretized time points, yielding $k$ archetypes for the functional observations and $c$ archetypes for the temporal dimension.

Each functional observation is approximated through convex combinations of the biarchetypes along both dimensions. The associated coefficient matrices $\boldsymbol{\alpha}_{n \times k}$  and $\boldsymbol{\gamma}_{c \times m}$ satisfy 
$\sum_{g=1}^k \alpha_{ig} = 1$, $\alpha_{ig} \geq 0$, $i=1,\ldots,n$, and
$\sum_{h=1}^c \gamma_{hj} = 1$, $\gamma_{hj} \geq 0$, $j=1,\ldots,m$.

Under these constraints, the residual sum of squares to be minimized is
\begin{equation}
\label{eq:biaa_fda}
\begin{split}
RSS
&= \left\lVert \mathbf{X} - \boldsymbol{\alpha}\mathbf{Z}\boldsymbol{\gamma} \right\rVert^2
 = \left\lVert \mathbf{X} - \boldsymbol{\alpha}\boldsymbol{\beta}\mathbf{X}\boldsymbol{\theta}\boldsymbol{\gamma} \right\rVert^2 \\
&= \sum_{i=1}^n \sum_{j=1}^m
\left(
x_i(t_j)
- \sum_{g=1}^k \sum_{h=1}^c
\alpha_{ig}
\left(
\sum_{l=1}^n \sum_{r=1}^m
\beta_{gl}\, x_l(t_r)\, \theta_{rh}
\right)
\gamma_{hj}
\right)^2 .
\end{split}
\end{equation}

In summary, $\boldsymbol{\beta}$ defines archetypes of functional observations, 
$\boldsymbol{\theta}$ defines arche\-types of temporal structure, while $\boldsymbol{\alpha}$ and $\boldsymbol{\gamma}$ describe how each function and each time point is expressed as a mixture of these archetypes.

 The biAA problem is solved using the algorithm of \cite{alcacerieee24}, available in both \textsf{R} and \textsf{Python}. The numbers of row and column archetypes, $k$ and $c$, are selected via an elbow-type criterion, choosing the pair $(k,c)$ at which the RSS surface flattens.

\section{Application} \label{res}

We are going to apply functional biAA to monthly time series of 10-year government bond yields of 23 European countries obtained from Eurostat: AT (Austria), BE (Belgium), CY (Cyprus), CZ (Czech Republic), DE (Germany), DK (Denmark), EL (Greece), ES (Spain), FI (Finland), FR (France), HU (Hungary), IE (Ireland), IT (Italy), LT (Lithuania), LU (Luxembourg), LV (Latvia), MT (Malta), NL (Netherlands), PL (Poland), PT (Portugal), SE (Sweden), SK (Slovakia), UK (United Kingdom). The sample period spans January 2001 to April 2025. This choice reflects data availability constraints: many countries lack observations prior to 2001, while for the United Kingdom data are not available beyond April 2025 at the time of the analysis. In addition, countries such as Bulgaria, Estonia, and Romania are excluded due to missing data within this period.

\subsection{Results}

Biarchetypes are calculated according to Section \ref{met}. The bidimensional screeplot shows that the best pair $(k,c)$ is $(3,3)$, resulting in $3$ archetypal countries and $3$ archetypal times.

\begin{figure}[t]
\centering
\includegraphics[width=0.46\textwidth]{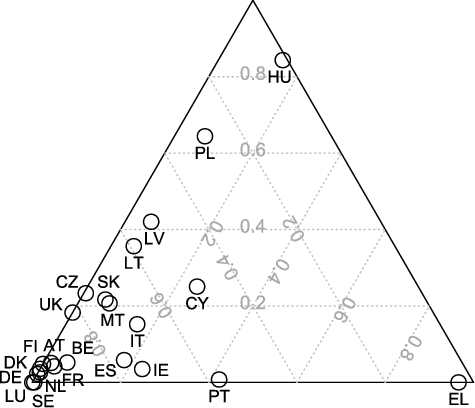}
\caption{Ternary plot of parameter $\boldsymbol{\alpha}$. The three archetypal countries are represented by the vertices of the triangle.}
\label{fig:tri}
\end{figure}

With respect to the archetypal countries, the proportion of each archetype attributed to each country approximation is reported in Table \ref{tab:arch} and illustrated in Figure \ref{fig:tri}. The first archetype is exemplified by Germany (DE), which is composed entirely of this archetype. Other economically advanced countries also exhibit very high shares of archetype 1, including Luxembourg (LU), Denmark (DK), the Netherlands (NL), Sweden (SE), Finland (FI), Austria (AT), France (FR), and Belgium (BE).

The second archetype is best represented by Greece (EL), which exhibits a share of $96.6\%$ of this archetype. Several Southern European countries also display substantial proportions of archetype 2, notably Portugal (PT), Cyprus (CY), Ireland (IE), Spain (ES), and Italy (IT); however, in all these cases, archetype 1 remains the dominant component.

Finally, the third archetype is exemplified by Hungary (HU), with $84\%$ of its composition corresponding to this archetype. Other countries (primarily located in Central Europe and the Baltic region) also exhibit notable shares of archetype 3, including Poland (PL), Latvia (LV), Lithuania (LT), Cyprus (CY), the Czech Republic (CZ), Slovakia (SK), Malta (MT), and the United Kingdom (UK). Among these countries, only Poland (PL) displays a majority share of archetype 3. All remaining countries are dominated by archetype 1, while Cyprus (CY) exhibits equal proportions of archetypes 2 and 3.

\begin{table}[]
\centering
\caption{Parameters $\boldsymbol{\alpha}$ multiplied by $100$. This shows the percentage of each archetypal country (1: red, 2: green, or 3: blue) in each country approximation.}
\label{tab:arch}
{\scriptsize
\begin{tabular}{c|rrrrrrrrrrrrrrrrrrrrrrr}
  & \textcolor[rgb]{1,0.018,0.055}{AT} & \textcolor[rgb]{1,0.060,0.058}{BE} & \textcolor[rgb]{1,0.495,0.500}{CY} & \textcolor[rgb]{1,0.006,0.306}{CZ} & \textcolor[rgb]{1,0,0}{DE} & \textcolor[rgb]{1,0,0.023}{DK} & \textcolor[rgb]{0.035,1,0}{EL} & \textcolor[rgb]{1,0.235,0.076}{ES} & \textcolor[rgb]{1,0,0.051}{FI} & \textcolor[rgb]{1,0.031,0.045}{FR} & \textcolor[rgb]{0.013,0.173,1}{HU} & \textcolor[rgb]{1,0.316,0.048}{IE} & \textcolor[rgb]{1,0.236,0.222}{IT} & \textcolor[rgb]{1,0.087,0.602}{LT} & \textcolor[rgb]{1,0.004,0}{LU} & \textcolor[rgb]{1,0.114,0.808}{LV} & \textcolor[rgb]{1,0.099,0.286}{MT} & \textcolor[rgb]{1,0.004,0.027}{NL} & \textcolor[rgb]{0.445,0.108,1}{PL} & \textcolor[rgb]{1,0.732,0.013}{PT} & \textcolor[rgb]{1,0,0.037}{SE} & \textcolor[rgb]{1,0.079,0.299}{SK} & \textcolor[rgb]{1,0,0.224}{UK} \\
\hline
\textcolor[rgb]{1,0,0}{1} & 93 & 90 & 50 & 76 & 100 & 98 & 3  & 76 & 95 & 93 & 1  & 73 & 69 & 59 & 99 & 52 & 72 & 96 & 29 & 57 & 96 & 72 & 82 \\
\textcolor[rgb]{0,1,0}{2} & 2  & 5  & 25 & 1  & 0   & 0    & 97 & 18 & 0    & 3 & 15 & 23 & 16 & 5  & 1  & 6  & 7  & 1  & 7  & 42 & 0    & 6  & 0    \\
\textcolor[rgb]{0,0,1}{3} & 5  & 5  & 25 & 23 & 0   & 2  & 0    & 6  & 5  & 4  & 84 & 4  & 15 & 36 & 0    & 42 & 21 & 3  & 64 & 1  & 4  & 22 & 18
\end{tabular}
}
\end{table}

\begin{figure}[t]
\centering
\includegraphics[width=0.94\textwidth]{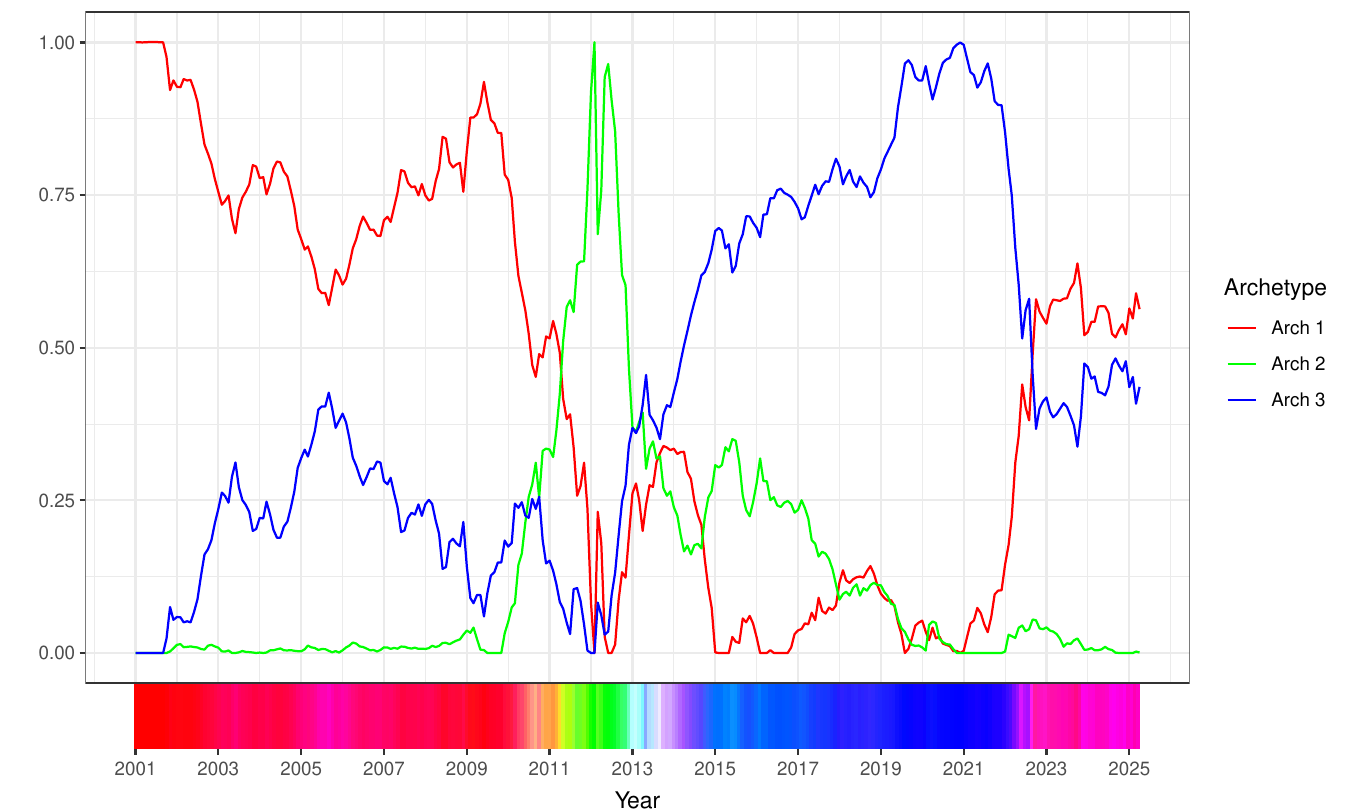}
\caption{Parameter $\boldsymbol{\gamma}$ over time, representing the mixture of archetypal times at each instant.}
\label{fig:timearch}
\end{figure}

Archetypal times are also identified. In an analogous manner to how each country is composed of contributions from each archetypal country, each point in time can be expressed as a combination of archetypal times. Figure \ref{fig:timearch} depicts the proportion of each archetypal time associated with each observation time, from January 2001 to  April 2025.

The first archetypal time is exemplified by the year 2001, which is characterized by the dot-com bubble crisis and high yields. This archetype remains dominant until April 2011. Although its influence subsequently declines as yields also decreases, it re-emerges as the dominant archetype from September 2022 onward, jointly with archetypal time 3.

The second archetypal time is represented by both February and June 2012, during which it reaches its maximum intensity. This period is associated with the euro area sovereign debt crisis, during which countries such as Greece (EL), Spain (ES), Italy (IT), Portugal (PT) and Ireland (IE) saw yields increase. Notably, in July 2012, Mario Draghi delivered his well-known statement that ``the ECB is ready to do whatever it takes to preserve the euro.'' This archetype dominates the time interval from May 2011 to March 2013.

The third archetypal time remains elevated throughout the COVID-19 crisis and reaches its peak in December 2020. In this period, yields were very low, even negative for countries such as Germany (DE) or Denmark (DK). Nevertheless, as previously noted, from September 2022 onwards, archetype 1 regains dominance, albeit with values that are only slightly higher than those of archetype 2. A similar coexistence of these two archetypes was also observed in 2005.

\section{Conclusions} \label{con}

We have proposed biarchetype analysis in the setting of univariate functional data for the first time. This unsupervised approach generalizes archetype analysis by jointly uncovering archetypal structures in both the observational units (countries, in our application) and the temporal domain. Specifically, both cases and time points are represented as convex combinations of biarchetypes, resulting in a parsimonious and highly interpretable summary of complex functional data.

We have illustrated the proposed methodology using 10-year government bond yields for European countries over the period 2001-2025. The analysis identifies three distinct temporal archetypes corresponding to the dot-com bubble crisis in 2001 (with high yields), the euro-area sovereign debt crisis (with increasing yields for some countries), and the COVID-19 crisis (with low yields). On the other hand, Germany, Greece, and Hungary have been revealed as country archetypes capturing yield dynamics across Europe. These results are consistent with those obtained using an alternative methodology based on multidimensional scaling of wavelet dissimilarities \cite{alcacer2026hplot}.

Finally, we believe that this methodology can be particularly valuable for the analysis of time series data, especially in economic and financial applications, where it may aid in the identification of business cycles.

\section*{Supplementary information} Code and data are available as supplementary information at {\url{https://epifanio.uji.es/RESEARCH/biaafda.zip}}.

\section*{Acknowledgement}
This work was partially supported by the Spanish Ministry of
Science and Innovation PID2023-153128NB-I00, PID2022-141699NB-I00, PID2020-118763GA-I00, and Generalitat Valenciana 
CIPROM/2023/66.

 \bibliographystyle{spmpsci} 
  \bibliography{refbiaa}

\end{document}